\documentclass[twocolumn,preprintnumbers,amsmath,amssymb]{revtex4-1}
\usepackage{epsfig}
\usepackage{color}
\usepackage{amsmath}
\usepackage{multirow}
\usepackage{graphicx}
\usepackage{subfigure}
\usepackage{floatrow}

\begin{document}

\title{Comparing machine learning potentials for water: \\ Kernel-based regression  and Behler-Parrinello neural networks}

\author{Pablo Montero de Hijes$^{*}$}
\affiliation{University of Vienna, Faculty of Physics, Kolingasse 14, A-1090
Vienna, Austria}
\author{Christoph Dellago}
\affiliation{University of Vienna, Faculty of Physics,  Kolingasse 14, A-1090
Vienna, Austria}
\author{Ryosuke Jinnouchi}
\affiliation{Toyota Central R$\&$D Labs., Inc., 41-1 Yokomichi, Nagakute, Aichi 480-1192, Japan}
\author{Bernhard Schmiedmayer}
\affiliation{University of Vienna, Faculty of Physics, Kolingasse 14, A-1090, 
Vienna, Austria}
\author{Georg Kresse}
\affiliation{University of Vienna, Faculty of Physics, Kolingasse 14, A-1090 Vienna, Austria}
\affiliation{VASP Software GmbH, Berggasse 21, A-1090
Vienna, Austria}

\begin{abstract}
 
In this paper we investigate the performance of different machine learning potentials (MLPs) 
	 in predicting key thermodynamic properties of water using RPBE+D3. 
	 Specifically, we scrutinize kernel-based regression and high-dimensional neural 
	 networks trained on a highly accurate dataset consisting of about 1,500 structures, 
	 as well as a smaller data set, about half the size, obtained using 
	 only on-the-fly learning. The study reveals that despite minor differences 
	 between the MLPs, their agreement on observables such as the diffusion constant 
	 and pair-correlation functions is excellent, especially for the large training dataset. 
	 Variations in the predicted density isobars, albeit somewhat larger, 
	 are also acceptable, particularly given the errors inherent to approximate density 
	 functional theory. Overall, the study emphasizes the relevance of the database 
	 over the fitting method. Finally, the study underscores the limitations of root mean 
	 square errors and the need for comprehensive testing, advocating the use of multiple 
	 MLPs for enhanced certainty, particularly when simulating complex thermodynamic 
	 properties that may not be fully captured by simpler tests.

\end{abstract}

\maketitle
$^*$pablo.montero.de.hijes@univie.ac.at

\section{Introduction}

  Computer simulations have significantly contributed to our understanding of water,
  starting in the
   late 1960s and early 1970s, when
   Barker and Watts through Monte Carlo simulations \cite{barker1969structure}
   and Rahman and Stillinger via molecular
    dynamics simulations \cite{rahman1971molecular} provided
     the first microscopic insights into water
      from a computational perspective.
      These pioneering studies were based on empirical potentials, 
 where the interactions are 
 designed to reproduce some physical
  behavior that is known in advance. 
  Electronic structure calculations for water were introduced in the 1990s \cite{laasonen1992water,laasonen1993ab,sprik1996ab,xantheas1995ab}
   and were gradually adopted \cite{grossman2004towards,chen2017ab,ruiz2018quest,gillan2016perspective,gaiduk2015density,schmidt2009isobaric,wang2011density,miceli2015isobaric,ceriotti2013nuclear,baer2011re,lin2012structure,vandevondele2005influence,zheng2018structural,zen2015ab}, 
   further increasing our understanding of the molecular mechanisms governing the behavior of water.
    However, first principles
     calculations were, and still are highly demanding computationally,
    becoming prohibitively expensive for
      large system sizes and long
       time scales. For this reason,
          empirical models 
        with different levels
         of approximation are still widely used and very popular
         \cite{molinero2009water,dyer2009site,jorgensen1981quantum,berendsen1981interaction,jorgensen1983comparison,berendsen1987missing,horn2004development,abascal2005potential,abascal2005general,izadi2014building,piana2015water,gonzalez2011flexible,habershon2009competing,stillinger1974improved,mahoney2000five,rick2004reoptimization,nada2016anisotropy,izadi2016accuracy,wang2014building,fuentes2014non,kiss2013systematic,reimers1982intermolecular,tainter2011robust,yu2003development,fanourgakis2006flexible,jiang2016hydrogen,pinnick2012predicting}.           
         In recent years, 
        machine learning based approaches
         have been established enabling us
          to perform calculations with first-principles
          accuracy but approaching the cost of much simpler force fields. \\

      In 1997, No {\em et al.} \cite{no1997description} described 
      the potential energy surface of the water dimer with  MP2 accuracy using a single feed-forward neural network. It took ten years, however, until 
      Behler and Parrinello cleared the path towards  
      high dimensional neural networks \cite{behler2007generalized} (BPNNP), which  were first 
       applied to the water dimer in 2012 \cite{morawietz2012neural}
       and to bulk water in 2016 \cite{morawietz2016van}. Subsequently,
         many other contributions for liquid water and ice followed \cite{singraber2018density,montero2023kinetics,cheng2016nuclear,kapil2016high,kapil2020inexpensive,cheng2019ab,morawietz2018interplay,reinhardt2021quantum,cheng2021phase,wohlfahrt2020ab}. Other machine learning methods 
       that have been applied to water 
       include permutation invariant
       polynomials (PIP) \cite{braams2009permutationally,yu2022q,zhu2023mb}, Gaussian processes regression based on kernel ridge regression (GAP) \cite{bartok2013machine}, support vector regression (SVR) \cite{bose2018machine}, deep neural network potentials (DP) \cite{zhang2021phase,torres2021using,lu202186,tisi2021heat,xu2023accurate,malosso2022viscosity,xu2020isotope,gartner2020signatures,gartner2022liquid,piaggi2022homogeneous}, Gaussian-moment neural networks\cite{P6274}, and, more recently, equivariant neural networks\cite{batzner20223,musaelian2023learning,mace,fu2022forces}.\\
       
       The selection and choice of a suitable machine learning approach is not always a simple matter. Generally, there are two key issues to address: accuracy --- how well does the machine learning approach approximate the ground truth, and speed -- how time-consuming are the calculations? Computational cost is trivial to address and easy to evaluate for different MLPs. It is much more difficult though to determine the accuracy of an MLP. 
       One very common and widely adopted approach is to compare to test data obtained using the first-principles approach. Commonly, the test data are of the same nature as the original training data (such as energies and forces), potentially obtained at different temperatures. However, this metric will only be reliable if the reference data covers all relevant regions of the
        potential energy surface. Furthermore, such a comparison does not consider whether the MLP yields stable trajectories during actual simulations. Finally, it is hard to determine what an acceptable accuracy is, how does a force error of say 30 meV/\AA\ translate into errors for physically relevant observables? Thus, the most meaningful approach to benchmark an MLP is
           to compare to physically relevant observables that are ideally also available from the first-principles calculations or experiments. This is one of the main goals of the present work.

Here, we focus on water at ambient conditions and two substantially different MLP approaches. 
In 2018, Nguyen {\em et al.} compared the representations of two-body and three-body interaction energies in small water clusters by means of GAP, BPNNP, and PIP finding similar accuracy \cite{nguyen2018comparison}.
       In the present work, we will push the comparison towards physically relevant observables. Specifically, we compare BPNNP with the machine learning force field (MLFF) approach implemented in VASP \cite{kresse1996efficient,kresse1996efficiency}. The MLPs in VASP are
        based on Gaussian process regression using Bayesian regression \cite{jinnouchi2019phase,jinnouchi2019fly,jinnouchi2020descriptors} or kernel-ridge regression. Independent of the type of the regression, similar two and three-body descriptors are used \cite{jinnouchi2020descriptors}. For brevity, we will denote them as kernel-based potentials (KbP).
      In particular, we compare the predicted pair-correlation functions and diffusion constants, as well as density isobars at zero pressure for a wide range of temperatures. The crucial question we pose is, whether the use of MLPs does introduce a significant error in the observables. We conclusively show that this is not the case, {\em i.e.}, the errors introduced by the two very different MLPs are virtually identical and very small for all calculated observables. This bodes well for the future use of MLPs in materials sciences and condensed matter physics. \\


         The remainder of this work is organized as follows.
          First, in Sec. \ref{sec:methods}, we explain how our datasets are produced. In Sec. \ref{sec:accuracy} the accuracy of training and computational cost of the models is studied. Then, in Sec. \ref{sec:results}
          we compare both methods for 
          actual observables, including
             the density maximum of water, the melting temperature,  the radial distribution functions, 
             and the self-diffusion coefficient. Finally, we
              highlight the main conclusions of this work in Sec. \ref{sec:conclusion}.\\

\section{Data acquisition \label{sec:methods}}

 The first part of this work consists of acquiring data from
  first principles calculations. For this purpose, we use
  projector-augmented-wave (PAW) potentials \cite{blochl1994projector,kresse1999ultrasoft} as implemented in VASP\cite{kresse1996efficient,kresse1996efficiency}. 
 Regarding the density functional, we select
  the RPBE \cite{hammer1999improved} +D3 \cite{grimme2010consistent} functional with standard damping (no Becke-Johnson damping, often referred to as zero damping). This has been shown to reproduce
   the behavior of water fairly well \cite{morawietz2016van}. 
  During this stage, all simulations are performed for 64 water
   molecules. To avoid errors, hard potentials (H$\_$h and O$\_$h) were used and the energy cutoff was set to 1100 eV. 
 The Brillouin zone was sampled at the $\Gamma$-point only. We also performed test calculations using more k-points, however, this changed the forces by less than 0.5~meV/\AA, which is entirely negligible.  
Despite this seemingly high cutoff, a Pulay stress of $\sim-0.75$ to $-0.85$ kbar is observed
 when comparing the pressure
 at 1100 eV with that at 2000 eV cutoff (that is, at 1100 eV, VASP will predict a volume that is too small). For water, this error is unacceptably large.
 To compensate for the basis set error introduced by the 1100 eV cutoff, we imposed 
    an external pressure of $-0.85$ kBar using the Parrinello-Rahman barostat during training (in VASP, the pressure convention is such that this corresponds to a tensile stress increasing the volume during the NPT simulations). A Langevin thermostat was used with a friction constant for
      both, the ionic and lattice degrees of freedom, set to 10 ps$^{-1}$.
   Moreover, a timestep of 1.5
fs was adopted to speed up the exploration of configuration space. If not noted otherwise, for production runs, the timestep was set to 0.5 fs. \\
 
The Bayesian regression as implemented in VASP was used for data acquisition at this stage.
For the on-the-fly MLPs a radial cutoff of 6~\AA\ and 8 basis functions were used for the pair descriptors, and for the three-body descriptors a cutoff of 4~\AA\ with 6 basis functions and a maximum angular $l$-quantum number of 2 was employed. These parameters were found to be optimal in an extensive hyperparameter search \cite{jinnouchi2023proton}.
The predicted Bayesian variance for the forces was used to decide whether first-principles calculations were performed or not. 
Based on this setup, a total of 473 first-principles calculations were performed and stored in a database.
The trajectory started by first heating cubic ice, Ic, until it melted at around 400 K, then equilibrating the structure at high temperature. Most training data were acquired during 
this initial phase briefly after melting. During cooling and annealing at various temperatures, very little additional data were added. Note that we cooled below the onset of vitrification several times to add low-temperature structures. The exact protocol including
 the selected Bayesian thresholds for the on-the-fly learning are compiled in Table  \ref{table:bayesian}. 
We note that the Bayesian threshold only correlates to the error that is related to the lack of data, and it does not
account for systematic errors caused by the model (e.g. finite interaction range or too few descriptors).
The required threshold needs to be continuously lowered as training continues
if additional structures in the training database are desired. However, increasing the number of training structures has no impact on systematic errors. This means that, at some point, adding more training structures does not improve the accuracy of the KbP noticeably.\\

Building on this preliminary KbP, we produce two different production-level datasets. First, to make a highly accurate dataset, we refitted the database using the 
fast kernel-ridge-based approach in VASP (forgoing the Bayesian regression) and
performed a parallel tempering (PT) simulation using VASP and this KbP. The PT run
was performed 
at an external pressure of $-0.85$ kBar and involved  20 replicas and a temperature range of 210--315 K.
From the trajectories, a total of 10,000 structures were selected. All of these structures were fed into VASP, the Bayesian error was evaluated and when above a threshold of 0.012 meV/\AA, a first-principles calculation was performed and the structure was added to the database.   
This increased the number of training structures to 1,053. The KbP was then updated, and the procedure was repeated a second time (including a second PT simulation), now using a Bayesian threshold of 0.01 meV/$\textup{~\AA}$ increasing the number of training structures to 1,495.
 Then, the energies, forces, and stress tensor for all the 1,495 structures were
  re-evaluated using a cutoff of 2000 eV producing our final first training set. Originally, we tried to avoid this last recalculation at an increased cutoff but we found that without re-calculation inconsistencies in the predicted volumes of about 2-3 $\%$ were observed
depending on whether the stress was fitted or not. We also note that the recalculation required less than one day on a 128 core AMD two-socket machine,
thus it adds very little extra cost and avoids inconsistencies 
caused by the Pulay stress.\\

\begin{table}[]
\caption{Temperature (T) protocol during the first stage of data acquisition. Also shown is
the total number of training structures. The three 200,000 step simulations are used
to make sure that no relevant "holes" are present in the potential energy surface. They add
very little training data. The sets marked with an asterisk are only part of the second training set. }
\label{table:bayesian}
\begin{ruledtabular}
\begin{tabular}{cccccccccccccccc}
Bayesian threshold [eV/\AA] &
T [K] & steps &  structures 
\\
\hline
  0.030  &  270-420 & 50,000 & 385  \\
  0.035 & 420-370 & 50,000 & 459  \\
  0.035 & 390-270 & 70,000 & 460  \\
  0.035 & 270-370 & 50,000 & 460   \\
 0.030 & 260-250   &200,000 & 463  \\
 0.035 & 350-350  & 200,000 & 469 \\
 0.030 & 275-235   & 200,000 & 473 \\  
 *0.025 & 350-250   & 200,000 & 523 \\
 *0.020  & 320-250  & 200,000 & 662 
\end{tabular}
\end{ruledtabular}
\end{table}

The second dataset also builds on the initial 473 structures. However, 
 we do not use parallel tempering for additional data acquisition. Instead, two additional temperature
  ramps with on-the-fly learning are performed. In the first one, 
  the Bayesian force-threshold 
  is decreased from 0.04 eV/$\textup{~\AA}$  to 0.025 eV/$\textup{~\AA}$ and the temperature is varied 
  between 350 K and 250 K. Then, by further decreasing the threshold to 0.02 eV/$\textup{~\AA}$
and performing a second ramp between 320 K and 250 K, we gathered in total 662 structures. Again, the dataset was
  post-processed with a plane-wave (PW) cutoff of 2000 eV to avoid Pulay stress errors. This smaller dataset allows us to assess whether on-the-fly learning is sufficient for data acquisition. \\

We stress that the setting we have used is of the highest possible quality. Neither an increase in the cutoff (2000 eV) nor an increase in the k-point set will change the values in our database noticeably. By iteratively refining the initial database through calculations in a production-like setting (parallel tempering), we maximize the dataset quality for the intended purpose. The only potentially debatable aspect is the cell size (64 molecules). However, given that the typical cell sizes are around 13~\AA\ during training --- more than double the cutoff we applied for the MLPs --- it seems improbable that finite size errors related to periodic boundary conditions can affect the MLPs.  
 In fact, it has been suggested recently that reference data
  from small water clusters are
   sufficient to train MLPs for bulk water~\cite{P5870,P6647,P6274}. \\

\begin{figure}[htb]
        \centering
 \includegraphics[width=3.1in]{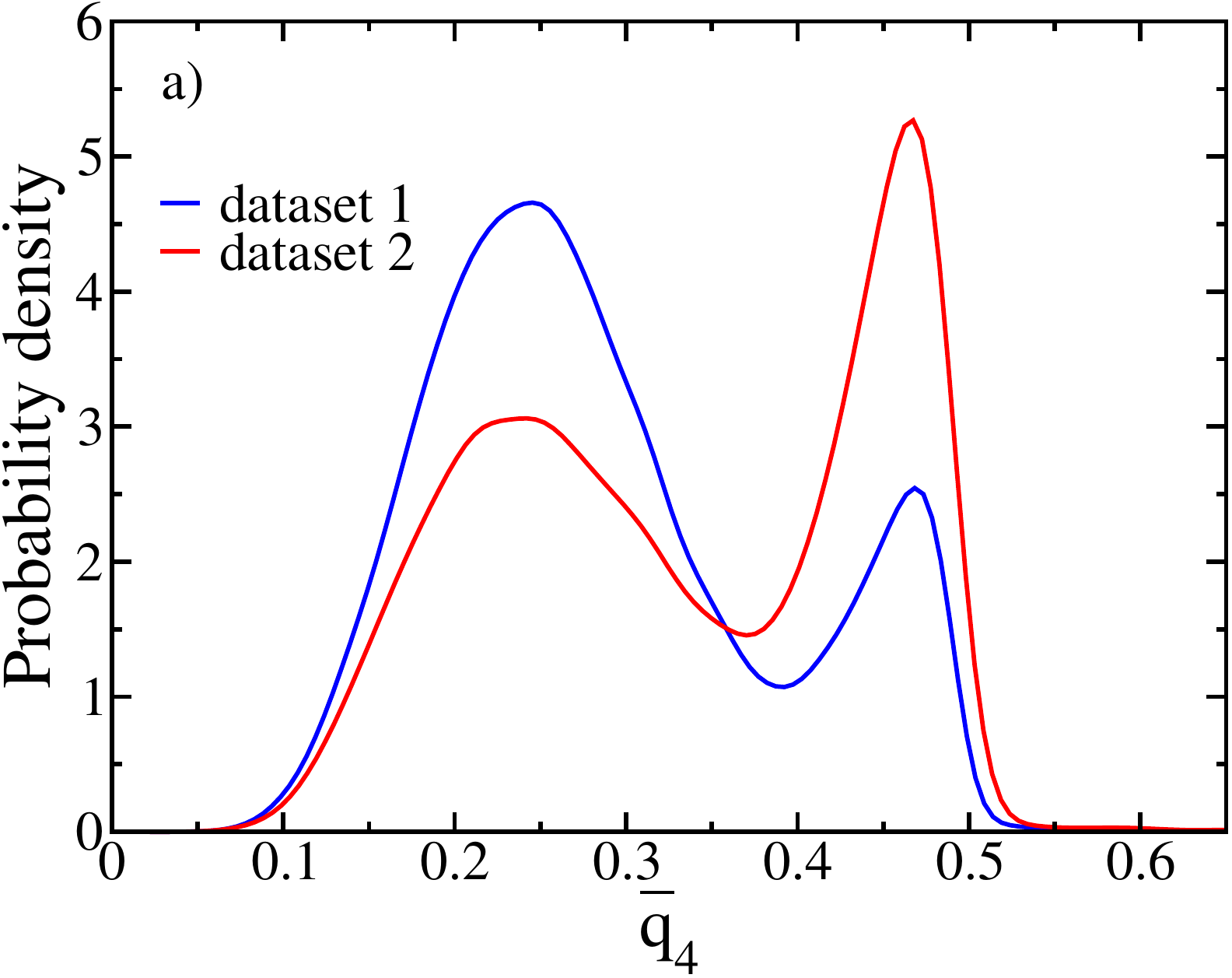}
        \centering
\includegraphics[width=3.1in]{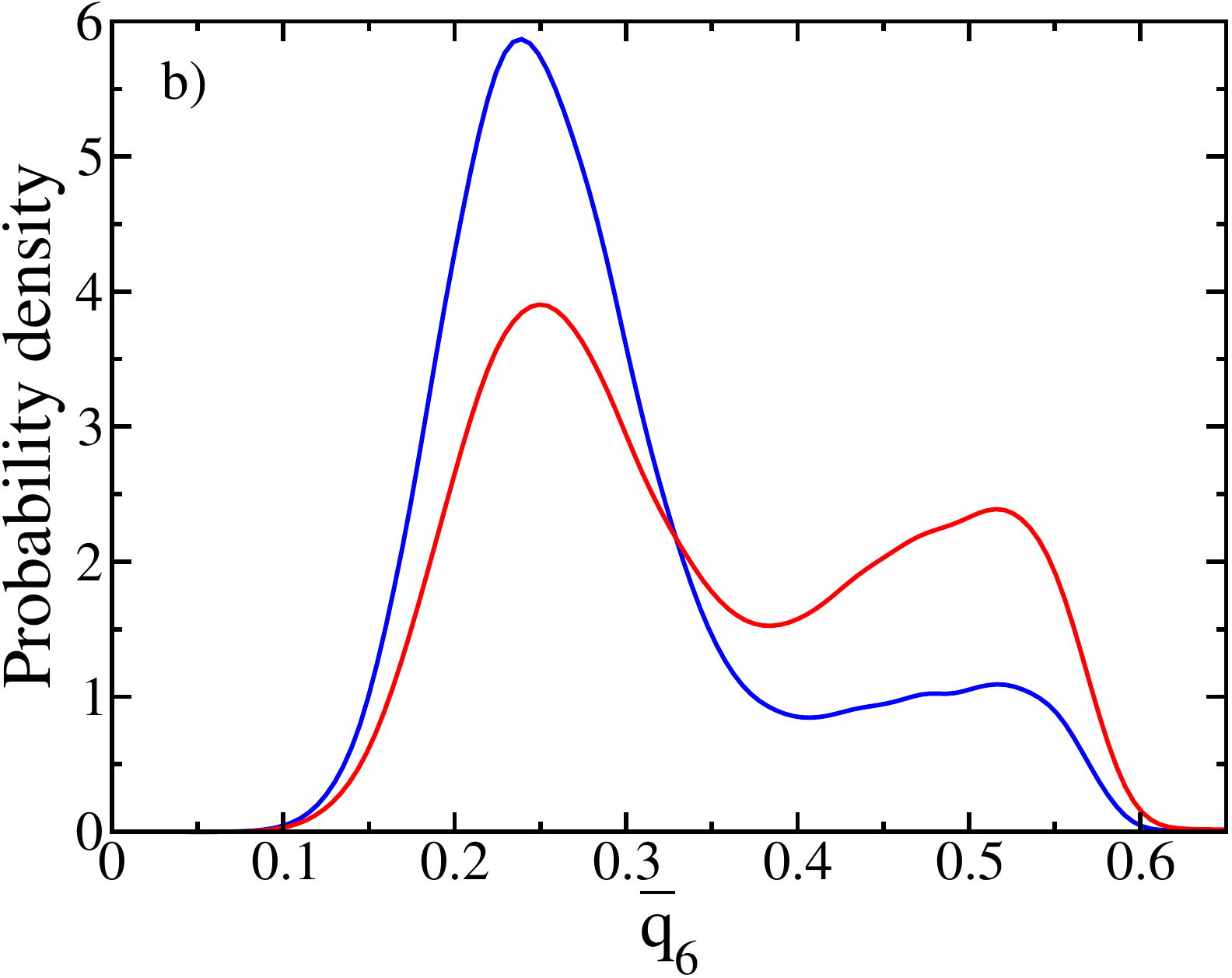}
\caption{\label{fig:kde_q46} Probability densities of averaged local bond-order parameters $\overline{q}_4$ (a) and $\overline{q}_6$ (b)
for dataset 1 (blue) and dataset 2 (red).  }
\end{figure}

A description of the datasets based on averaged  Steinhardt local bond order parameters
\cite{lechner2008accurate},  $\overline{q}_4$  and $\overline{q}_6$ with 3.5 $\textup{~\AA}$ cutoff,  is shown 
 in Fig. \ref{fig:kde_q46}. There, the probability densities are shown,
  where the left peak corresponds
  to liquid-like structures and the right one to ice Ic\cite{sanz2013homogeneous}.
  In the case of the larger dataset, the global distribution is
    strongly biased toward liquid-like structures. However, in 
    the smaller dataset, 
     there are more ice Ic structures than
     liquid ones according 
    to $\overline{q}_4$, whereas  $\overline{q}_6$ suggests the opposite. 
     Therefore, the smaller dataset is rather equally distributed between 
     ice Ic, partially melted ice Ic, and liquid water. Although
       liquid structures already contain the reference building blocks for
       modeling ice phases through BPNNP \cite{monserrat2020liquid,guidarelli2023neural}, this does not apply the other way around. Thus, the second dataset contains,
        in principle, fewer relevant structures.

\section{Accuracy and performance  \label{sec:accuracy}}

 First, we compare the accuracy
  of our models by using the standard procedure, i.e., by computing the 
  root mean square error (RMSE) 
  for forces, energies, and in the case
   of the KbP also the stress tensor.
 We note that 
   this sort of metric is sensitive
    to different factors. For example, a reduced RMSE does not necessarily indicate a universally superior model. If the database contains correlated data, the RMSE might decrease due to highly correlated data but this tells little about the stability of the MLFF. Consequently, one could obtain a smaller error even when the model is not reliable and robust. In essence, as the model has to tackle more of the relevant configurational space, the challenge increases, and the RMSE may rise despite the model becoming more robust.
           As explained above,  
       we have produced two different
        datasets with varying
         sizes and data acquisition
          approaches. The differences
           in the RMSE of energy and forces of these sets are small, 
            but nevertheless, we will see that stability issues appear later
            when observables are
            computed.\\
 
\begin{table}[htb]
\caption{Comparison of the accuracy of the KbP and BPNNP models trained 
 on the highly accurate dataset 1 and on the on-the-fly dataset 2.
  The root mean square errors (RMSE)  are given for (from left to right)
   energy  (meV/atom), forces  (meV/$\text{\AA}$), and stress tensor (kbar). 
  }
\label{table:acc}
\begin{ruledtabular}
\begin{tabular}{cccccccccccccccc}
method/software &
dataset & E$_{\rm RMSE}$  & F$_{\rm RMSE}$  & S$_{\rm RMSE}$\\
\hline
\multirow{2}{3cm}[5mm ]{} & 1 & 0.30 & 27 & 0.16\\ 
\multirow{2}{2cm}[4mm ]{KbP/VASP} & 2 & 0.30 & 27 & 0.16 \\ 


\hline
\multirow{2}{3cm}[5mm ]{} & 1 & 0.26 & 28 &  - \\ 
\multirow{2}{2cm}[4mm ]{BPNNP/n2p2} & 2 & 0.30 & 27 & - \\  

\end{tabular}
\end{ruledtabular}
\end{table}

Results of this error analysis are shown in Table  \ref{table:acc}.
The root mean square  error in the KbP approach 
is  $\sim$27  meV/\text{\AA} 
 in the force components,  $\sim$0.30   meV/atom in the energy, and  $\sim$0.16 kBar in the stress tensor for both data sets.
 As an additional test, we picked 40 configurations with 128 molecules from two KbP parallel tempering (PT) simulations
(see Sec. \ref{sec:resultsdensity}). These span a temperature range from 210 to 330 K.
For these structures, first-principles calculations were performed and compared to the KbP for the large training set. 
The average errors are in line with the 64 molecule training set error, e.g. 27 meV/\AA\ for the force error, 0.2 meV/atom for energies, and 0.1 kbar (errors per atom and pressure errors decrease like one over the square root of the number of atoms, if there are no relevant long-range correlation effects). This suggests that finite-size errors related to the training set can be neglected.

 \begin{figure}[htb]
\centering
\includegraphics[width=3.1in]{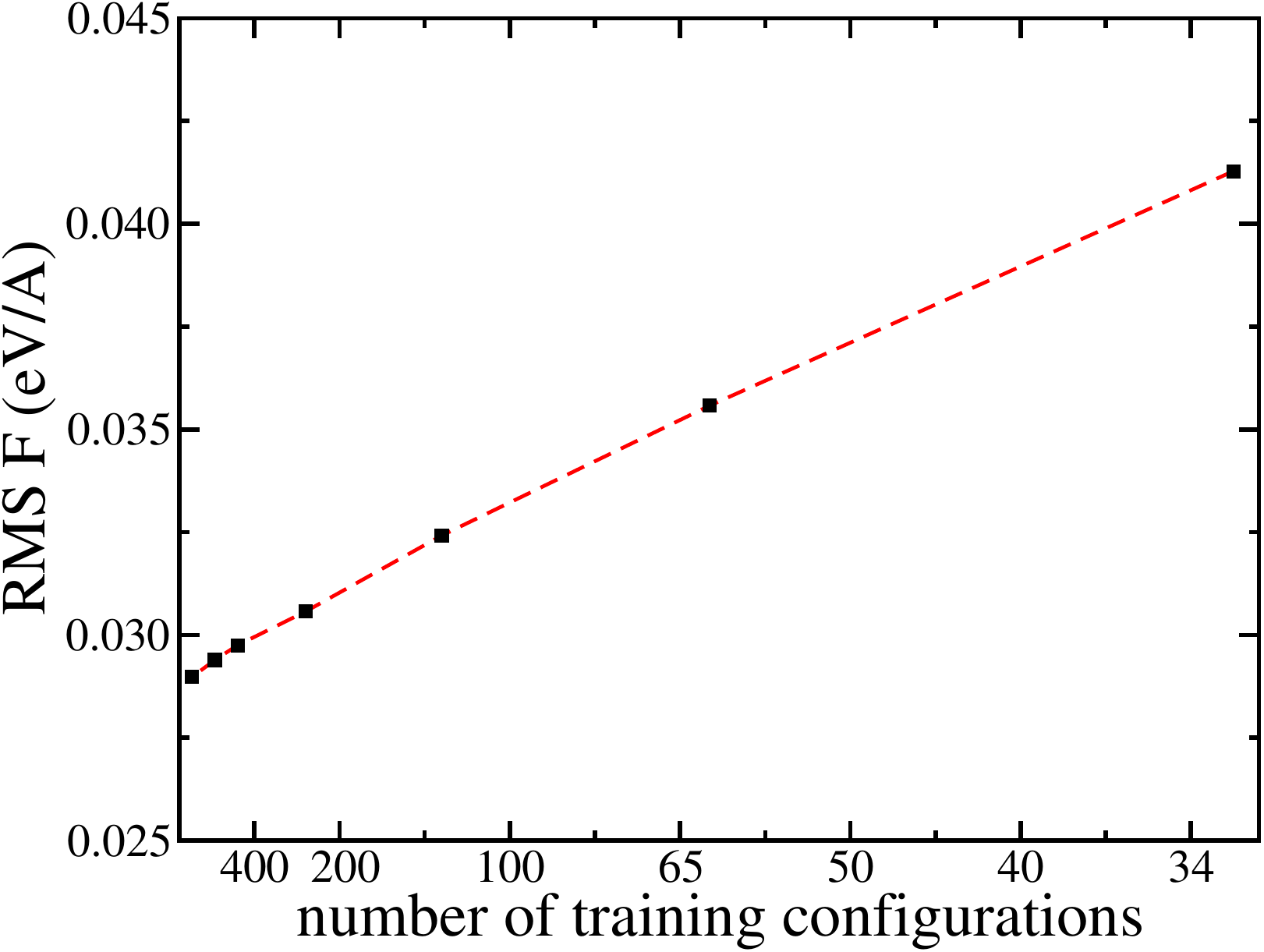} 

\caption{\label{fig:learning}  Learning curve for the KbP approach, trained on a subset of the large data set and evaluated on an independent test set (see text for details). As the error does not approach 0, log-log plots do not convey a concise message and therefore we chose a reciprocal representation for the $x$-axis. The error appears to be inversely proportional to the number of training structures.}

\end{figure}

In Fig. \ref{fig:learning}, learning curves are shown for the KbP. Here we have used subsets of the large training set, ranging from 32 structures up to the full dataset. 
The force errors were evaluated for the about 200 structures only present in the small data set (marked with ``*'' in Table  \ref{table:bayesian}). We see fairly fast convergence with respect to the number of training structures with diminishing improvements above about 500 structures. However, we also see that the learning curve levels off at about 28 meV/\AA\ (the 200 test structures contain only a few ice-like structures and relatively more high-temperature structures than the training set).

  
Regarding the  training of the 
Behler-Parrinello-Neural-Network (BPNNP) model, the network architecture consists of 2 hidden layers with 25 nodes each. Softplus is used as activation function and the RMSE as loss function. The cutoff is
 set to 6.36 \AA . The symmetry functions are
 the same as in Morawietz at al. \cite{morawietz2016van}
 leading to 2827 parameters (weights and biases) per BPNNP.
We train and simulate
  our BPNNP  with the n2p2 interface \cite{singraber2019parallel} for LAMMPS \cite{thompson2022lammps}. 
  In particular, 
  four different BPNNPs were trained
  on each of the two datasets. Again, we find negligible differences
   in the  RMSE for forces and energies for a given dataset.
   However, small differences appear between the two sets. The first one 
    results in better accuracy for
    the energy but it is slightly worse for the forces than the second. 
 The tiny differences between the two datasets are not entirely surprising. The small difference between the BPNNP and KbP approach is, however, to some extent remarkable. As a matter of fact, both approaches are based on two- and three-body correlation functions and similar radial cutoffs. They also share a lack of explicit modeling of four-body correlations and long-range electrostatic effects. \\

\begin{figure}[htb]
\centering
\includegraphics[width=3.1in]{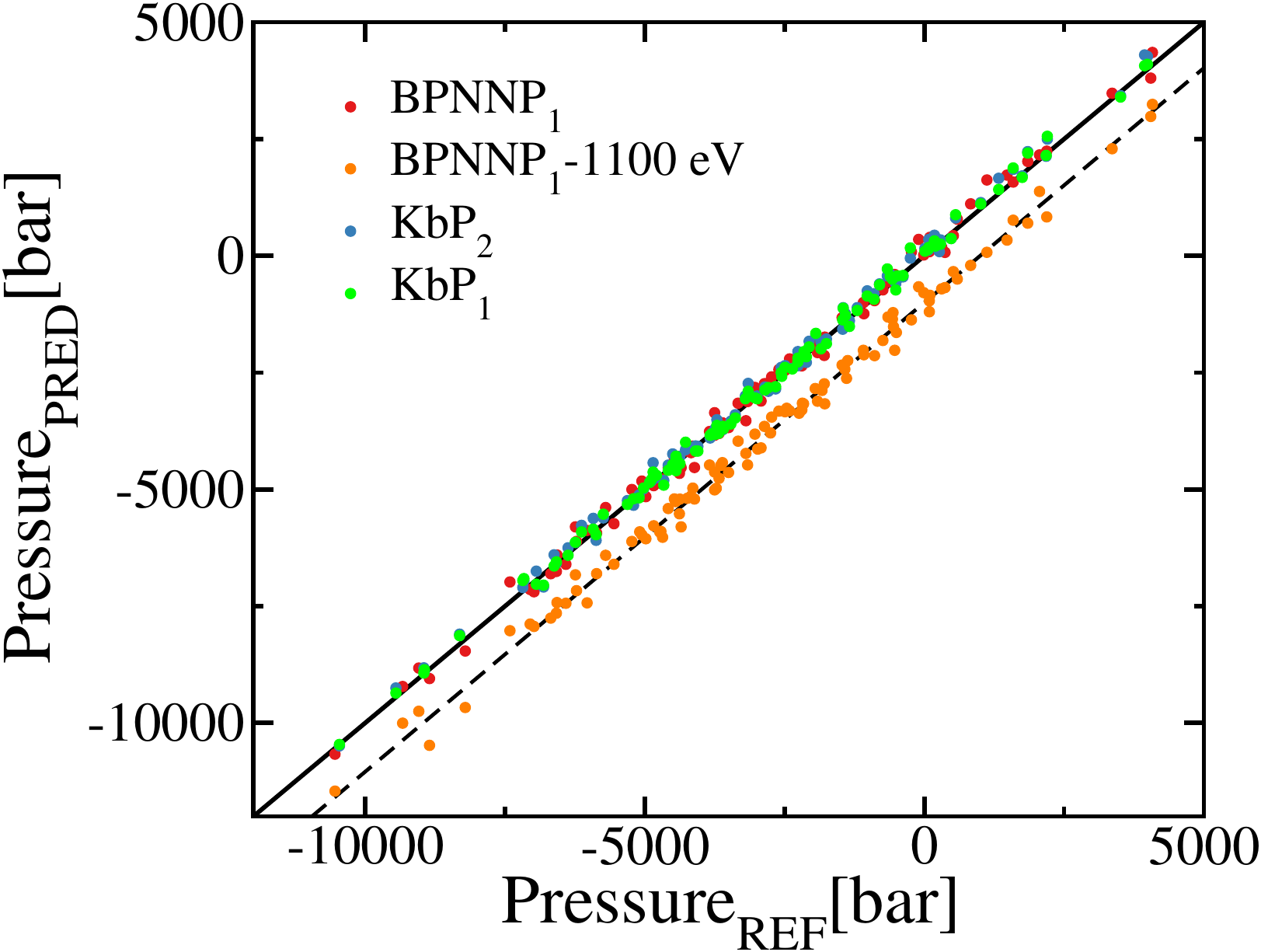} 

\caption{\label{fig:press}  Predicted (PRED) vs. reference (REF) pressure for
KbP, which learns the stress tensor, and BPNNP, which does not learn it.
 Also shown are results obtained with a BPNNP trained on a database without
  re-calculation at 2000 eV, i.e. trained on the original 1100 eV cutoff dataset.
 The offset in this case is  $-1$~kbar, which is not in full agreement with the expected $\approx -0.8$~kbar.  The coefficients (c$_0$, c$_1$) of the linear
  regression  $y=c_0 + c_1\cdot x$  
   are (0.051 kbar, 1.010) for
    BPNNP$_1$, ($-$1.00 kbar, 0.997) for
    BPNNP$_1$-1100 eV, (0.066 kbar, 1.007)
    for KbP$_1$, and (0.102 kbar, 1.007) for KbP$_2$.}
\end{figure}

At this point, we can conclude that the two approaches yield overall similar accuracy in the standard evaluation procedures.
    Since the BPNNP cannot learn the stress tensor whereas the KbP does, 
    it is useful to predict the pressure for test structures using the BPNNP.
     The results are shown in Fig. \ref{fig:press}. Clearly, the BPNNP is capable
      of providing a good pressure prediction through the learning of forces and energies without
       explicitly learning the stress tensor. Interestingly, when the training set data 
        is not re-calculated at 2000 eV, that is, when we use the initial data at 1100 eV, the average deviation in pressure is about $-1$~kbar.
        The first-principles calculations however indicate a difference of $-0.75$ to $-0.85$ kBar between 1100 eV and 2000 eV. This discrepancy relates to the already mentioned issue that results are less robust and reliable if the original on-the-fly database is learned. We add here that during molecular dynamics simulations, VASP does not change the basis set. As the volume fluctuates, this causes slight changes in the effective plane wave cutoff, and this in turn slightly affects the energies and the stress tensor. This problem is eliminated by recalculating the entire database at the end. We emphasize that this problem is specific to molecular liquids, as we have never observed differences between learning the on-the-fly database or a recalculated database for solid-state systems.\\

Let us now assess the computational cost associated with each machine-learning method
 for the same computational resources. In both cases, the machine learning
  codes are optimized to run on CPUs. 
  Specifically, we use a dual-socket AMD 7713
node (64 cores per socket). We cover system sizes from 384 atoms up to 384,000 atoms and run the simulations in the NVT ensemble. 
The summary of the computational cost is shown in Table \ref{table:eff}. For VASP, we note that these calculations were performed in the kernel-based fast mode using ridge regression, forgoing Bayesian regression. No uncertainty predictions are possible in this mode, but the computational time is reduced by roughly a factor of 30. 
As can be seen in the table,
 the total run time (wall clock) is slightly lower for the BPNNP. We note that the  KbP shows improved timings when reducing the number of cores for few atoms, whereas
   the BPNNP yields the best timings for the largest number of cores independently of
    the system size. We believe this is related to the huge bandwidth requirements for the evaluation of the kernels. VASP allows to sparsify the angular descriptors \cite{jinnouchi2020descriptors}.
 This reduces the bandwidth requirements, improves the strong scaling, and is particularly beneficial for small systems, as shown by the numbers in parenthesis for KbP.

\begin{table}[htb]
\caption{Computational cost of energy and force prediction using a dual socket AMD 7713 node (64 core/socket).
 Smaller systems optimally require less cores for the KbP running VASP, whereas
  for the BPNNP running with n2p2 in LAMMPS using all cores provides better performance. Values in parentheses for KbP use only 30~\% of the angular descriptors and 64 cores (128 cores for 24,576  and 384,000 atoms). Compute time depends to some extent on compilers, libraries, and MPI versions.}
\label{table:eff}
\begin{ruledtabular}
\begin{tabular}{cccccccccccccccc}
method/software &
atoms & cores &
$\mu$s/(atom$\cdot$step)
\\
\hline
\multirow{2}{3cm}[5mm ]{} & 384 & 128 & 9.2 \\
\multirow{2}{3cm}[5mm ]{} & 3,072 & 128 & 4.6  \\ 
\multirow{2}{2cm}[5mm ]{BPNNP/n2p2} & 24,576 & 128 & 4.1 \\  
\multirow{2}{3cm}[5mm ]{} & 384,000 & 128 & 4.7 \\
\hline
\multirow{2}{3cm}[5mm ]{} & 384 & 16 & 13.7 (3.8) \\
\multirow{2}{3cm}[5mm ]{} & 3,072 & 64 &7.2 (4.6)\\ 
\multirow{2}{2cm}[5mm ]{KbP/VASP} & 24,576 & 128 & 5.1 (4.2) \\ 
\multirow{2}{3cm}[5mm ]{} & 384,000 & 128 & 5.3 (4.7) \\


\end{tabular}
\end{ruledtabular}
\end{table}

We have also tested dataset 1 with the recent machine learning approach {\em Allegro}, a local equivariant deep neural network interatomic potential\cite{musaelian2023learning}. A  comparison of different methods is not straightforward in this case as {\em Allegro} is optimized
   to run on GPUs. Using a \textit{single} Nvidia A100 GPU, an {\em Allegro} model with comparable
     accuracy to our models, unfortunately, requires 10 times longer than the 2 socket AMD nodes using KbP and BPNNP. In addition, {\em Allegro} can not yet predict the stress tensor and the use of 
     NPT simulations is still
      under development. Therefore we have not included the {\em Allegro} model 
      in further comparisons. 
     Since better scaling with increasing GPU resources is expected for {\em Allegro}, 
     the optimal ML approach for a given application may still depend on the available resources and the number of atoms in the simulation.


\section{Physical properties  \label{sec:results}}

 In this section, we apply both of our machine-learning potentials to obtain
  structural, thermodynamic, and dynamical properties of water with first-principles accuracy. 
 We start by inspecting the density maximum and the melting temperature, 
     followed by the partial radial distribution functions, and finally
      the self-diffusion coefficient. For compactness, we use the notation BPNNP$_1$ and KbP$_1$ for MLPs trained on the first larger dataset and BPNNP$_2$ and KbP$_2$ for MLPs trained on the second smaller on-the-fly dataset.

\subsection{Density maximum and melting temperature} \label{sec:resultsdensity}

 To determine the density maximum of liquid water, we performed PT simulations for 128 molecules with temperatures ranging 
 from 200 K up to 335 K at 1 bar. For all BPNNP, 
 the duration of the simulations 
 is 1 ns in each of the 32 replicas (2 million steps each). Four PT runs are launched for each of the four 
  BPNNP that were trained for each of the two training sets. The first half of each trajectory is discarded as
   equilibration. In fact,  
   we find that some lower-temperature replicas get kinetically trapped due to vitrification and stop to exchange
 Therefore, values for temperatures below 227 K are not shown in the plots.
  In the case of the KbP, we used only 20 replicas spanning a temperature range between 210 and 330 K. The spacing was roughly exponential. To make the simulations using the KbP more effective, the mass of hydrogen was increased to 8 a.m.u. and the timestep was set to 1.5 fs. Here only a single run was performed but using 10 million time steps giving a total simulation time of 15 ns for each replica. Additionally, a 7 ns parallel tempering run was performed for equilibration. As for training, Langevin thermostats were used for the atoms and the cell volume, but with a friction coefficient of 2 ps$^{-1}$. Likely because of the use of Langevin thermostats, we require longer simulation times to obtain good statistics in VASP.\\

 \begin{figure}[htb]
  
\centering
  \includegraphics[width=2.8in]{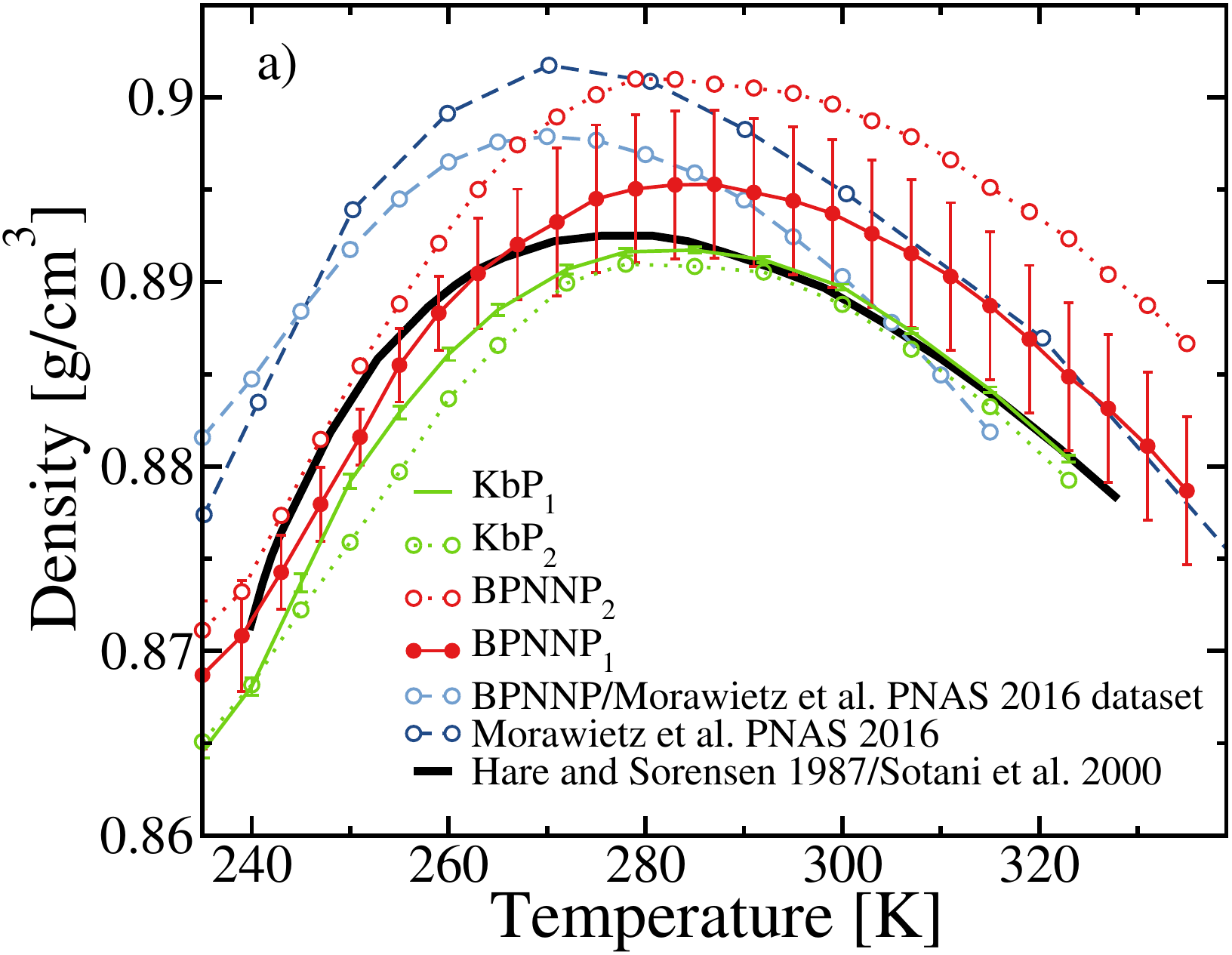} 
  \includegraphics[width=2.8in]{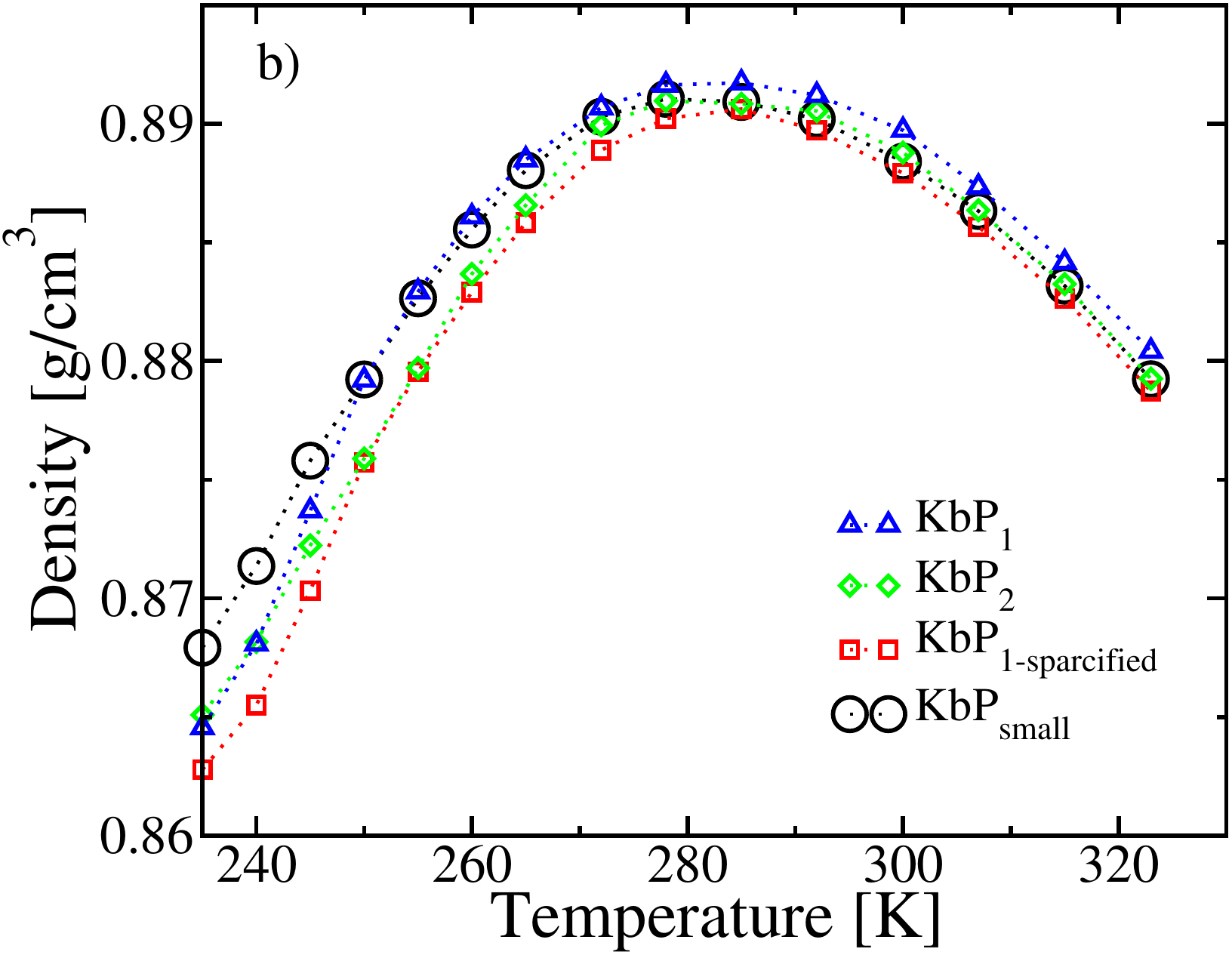}
\includegraphics[width=2.8in]{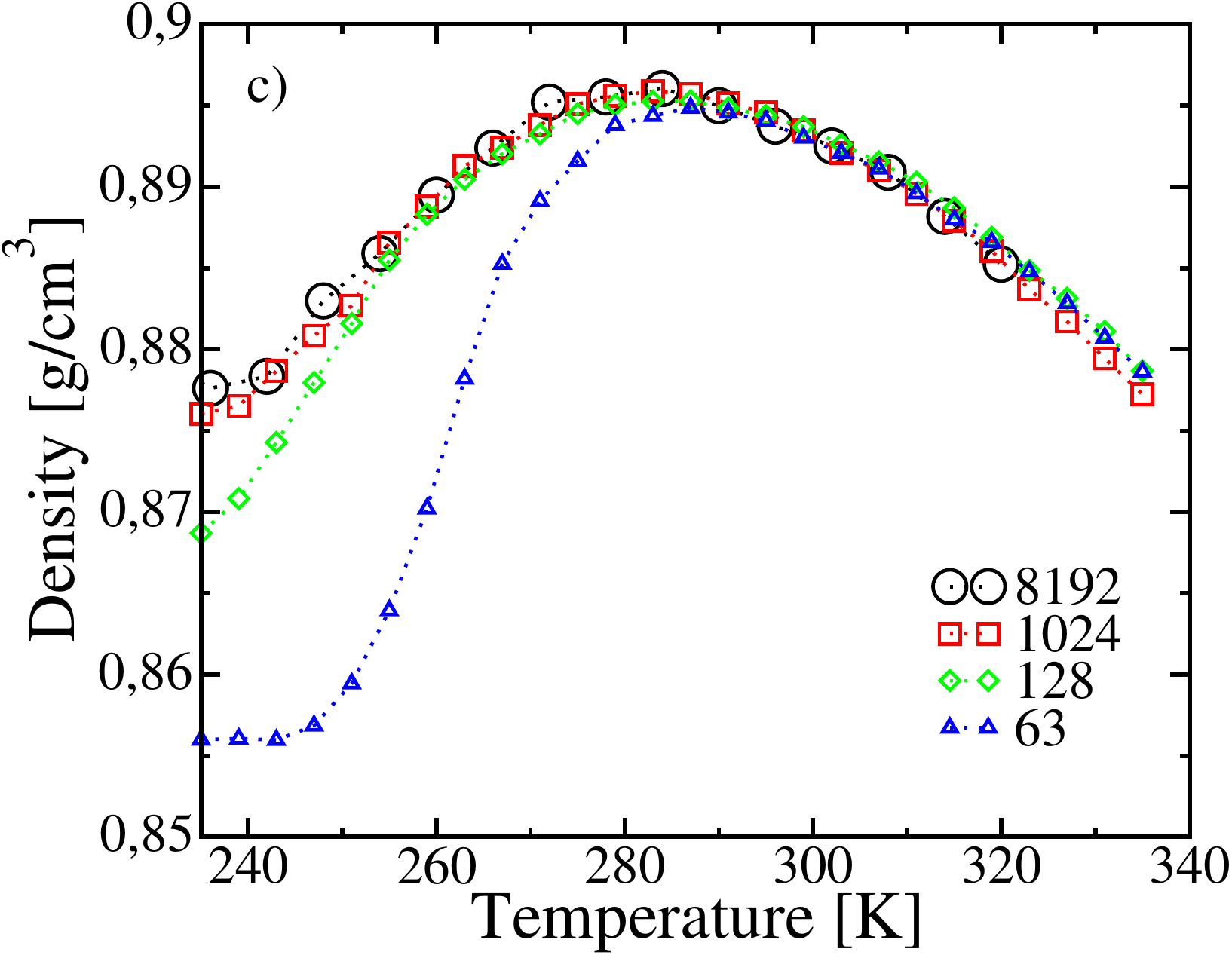} 
\caption{\label{fig:maxd} a) Density as a function of temperature at a pressure of 1 bar. 
The error bars for BPNNP$_1$ (solid red circles) are determined 
 from the standard deviation over the four different BPNNP trained on 
 dataset 1. For BPNNP$_2$ (empty red circles) only one run was
  stable and therefore no error bars are shown. KbP results are shown in green (solid for dataset 1 and
   empty for dataset 2); error bars for kbP$_1$ are determined using block averaging. The solid black line corresponds
   to experimental measurements \cite{hare1987density, sotani2000volumetric,holten2012thermodynamics} and has been shifted down by 0.1075 g/cm$^3$ to facilitate the comparison. Empty dark blue 
    circles show results from  Morawietz {\em et al.} \cite{morawietz2016van}. Empty light blue circles correspond to the result of a parallel tempering
     simulation for a BPNNP trained here on the
      dataset of  Morawietz {\em et al.} \cite{morawietz2016van}. b) Comparison
       of various KbPs, including one trained only on 250 structures and one where 
          the descriptors have been sparsified. c) Density versus temperature for different system sizes ranging from 63 to 8,192 molecules computed on the same
           BPNNP.  }
\end{figure}

The results of these PT simulations are shown in Fig. \ref{fig:maxd}.
 In the case of the BPNNP for the small second training set we encountered some issues: for three out of four 
    trained potentials the simulations became unstable, 
    and only for one, the four trajectories were completed without errors. Note
     that our RMSEs were almost the same as those for
      the BPNNP trained on the first, larger training set. This
     instability was never observed for the KbP trained on the same data,
      suggesting that BPNNPs require larger input data sets to become stable and robust. 
      Moreover, we find an intrinsic error of about 1$\%$  in the density
       as obtained from the different runs on BPNNP as indicated by the error bars. 
       This sort of uncertainty cannot be assessed for a KbP, because it is largely
        a deterministic approach. In any case, the
         difference between KbP$_1$ and KbP$_2$
          which were trained on different
          datasets is very small.

    Also shown in Fig. \ref{fig:maxd} are
     the results of Morawietz {\em et al.} \cite{morawietz2016van} who used a BPNNP trained on data obtained using the same density functional but computed with FHI aims. This dataset was also refitted here, and parallel tempering simulations analogous to the other BPNNPs were performed. The difference to the original publication is about of the same order as the discrepancies between different BPNNPs fitted to the datasets generated in the present work. 
      As can be seen in the figure, our results are in fairly good agreement with the previous results, for the present training data, however, the maximum is shifted to slightly
     higher temperatures. 
       We find the density maximum for the RPBE+D3 functional at approximately
        284 K using the BPNNP and 283 K using the KbP. We also compute
         the melting temperature. First, by
          running several trajectories for about 10 ns in the
           anisotropic NpT ensemble at
            standard pressure and different 
             temperatures around the expected melting.
             Once, we find two consecutive
              temperatures in which ice barely
               grows or melts, we extract one 
               of these configurations and continue
                the run in the NpH ensemble
                 until the system  converges to the
                  melting temperature. For
                   the BPNNP, the value
                    obtained is 277 $\pm$ 1 K and
                     for KbP 279 $\pm$ 1 K. Therefore,
                     the difference between
                      the melting and density maximum temperatures are 7 °C
                      for the BPNNP and 4°C for
                       the KbP, being in
                        good agreement with experiments,
                        where value for the
         temperature of maximum density is 4°C.
        Although
        the RPBE+D3 functional underestimates the density giving almost 10$\%$ smaller densities, the theoretical and experimental curves are
         very similar as shown in Fig. \ref{fig:maxd}, where
          the experimental curve \cite{hare1987density, sotani2000volumetric,holten2012thermodynamics} is shifted down by 0.1075 g/cm$^3$. The shape
        of the maximum in the density 
         is well reproduced.

         Two more results for KbP are shown in Fig. \ref{fig:maxd} b) to assess whether further computational savings are possible without compromising quality. Both force fields result in force errors around 31 meV/\AA, only slightly larger than for the other MLPs. The circles  indicate results obtained using only 250 training structures from the large training set (compare also Fig. \ref{fig:learning}).  Although the results are still very good, we see an increase in the density at low temperatures, which might be due to an insufficient number of training structures in the strongly
         supercooled regime. Furthermore, we report results for a KbP using only 30~\% of the angular descriptors (compare Table  \ref{table:eff}), which speeds up the calculations by up to a factor of 2. Results are again in very good agreement with the other KbPs. Inspecting the four different curves for the KbP it is hard to make out systematic trends. It rather seems that these are similar in nature, but smaller than the variations for the BPNNPs, and hence a measure of the residual errors of the force fields. We note that the variations increase towards the supercooled liquid state, which might be related to a lack of training structures and thus increased uncertainty in that regime.

         In the present work, we have also investigated how different system sizes affect the results. We have restricted this analysis to the BPNNPs, but expect similar results for the KbP. Figure \ref{fig:maxd} c) clearly indicates that 128 molecules are required to obtain technically converged results. Vitrification occurs very often in the smaller ensembles with 63 molecules. We also see that there are considerable system size dependencies below 250 K, where glassy states set in. This likely indicates that our data below 250 K should be considered with some caution and would require a more careful system size analysis, ideally with longer equilibration. Around the density maximum though the size effects are insignificant if at least 128 molecules are used.

\subsection{Radial distribution functions }

\begin{figure*}[htb]

\centering

\begin{picture}(510,150)
 \put(0,0){\includegraphics[width=2.15in]{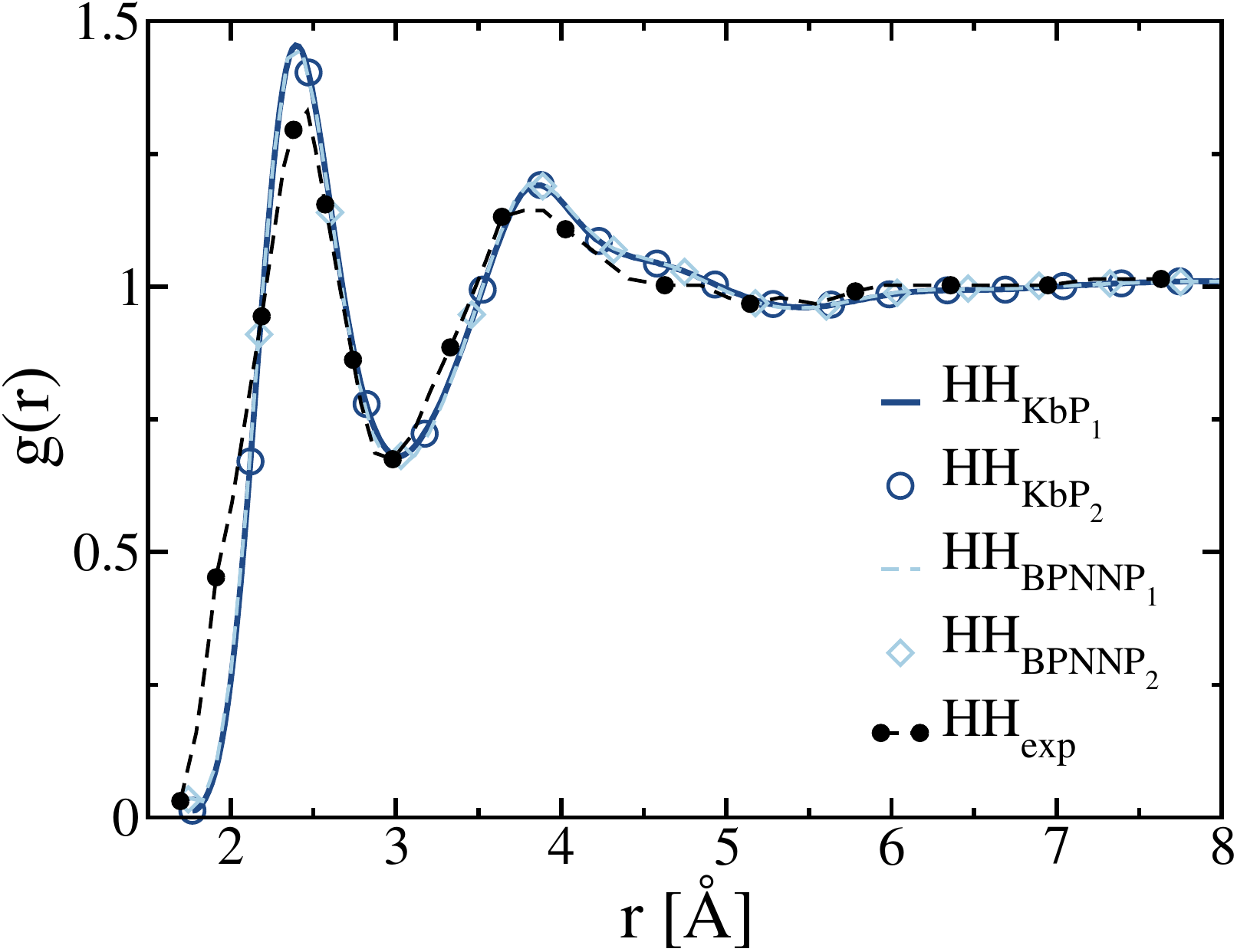} }
    \put(0,130){a)}
\put(170,0){\includegraphics[width=2.15in]{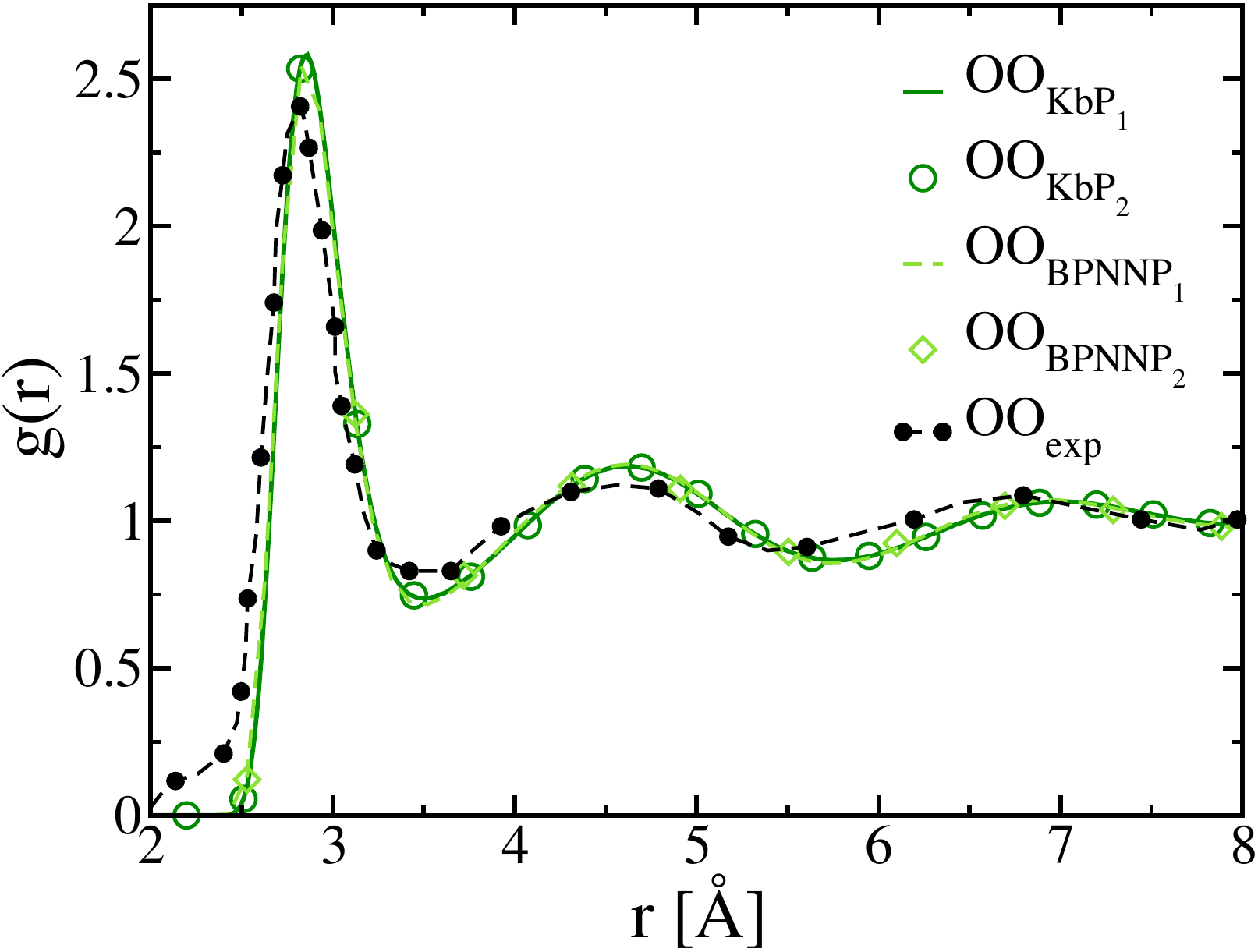} }
 \put(170,130){b)}
\put(340,0){\includegraphics[width=2.15in]{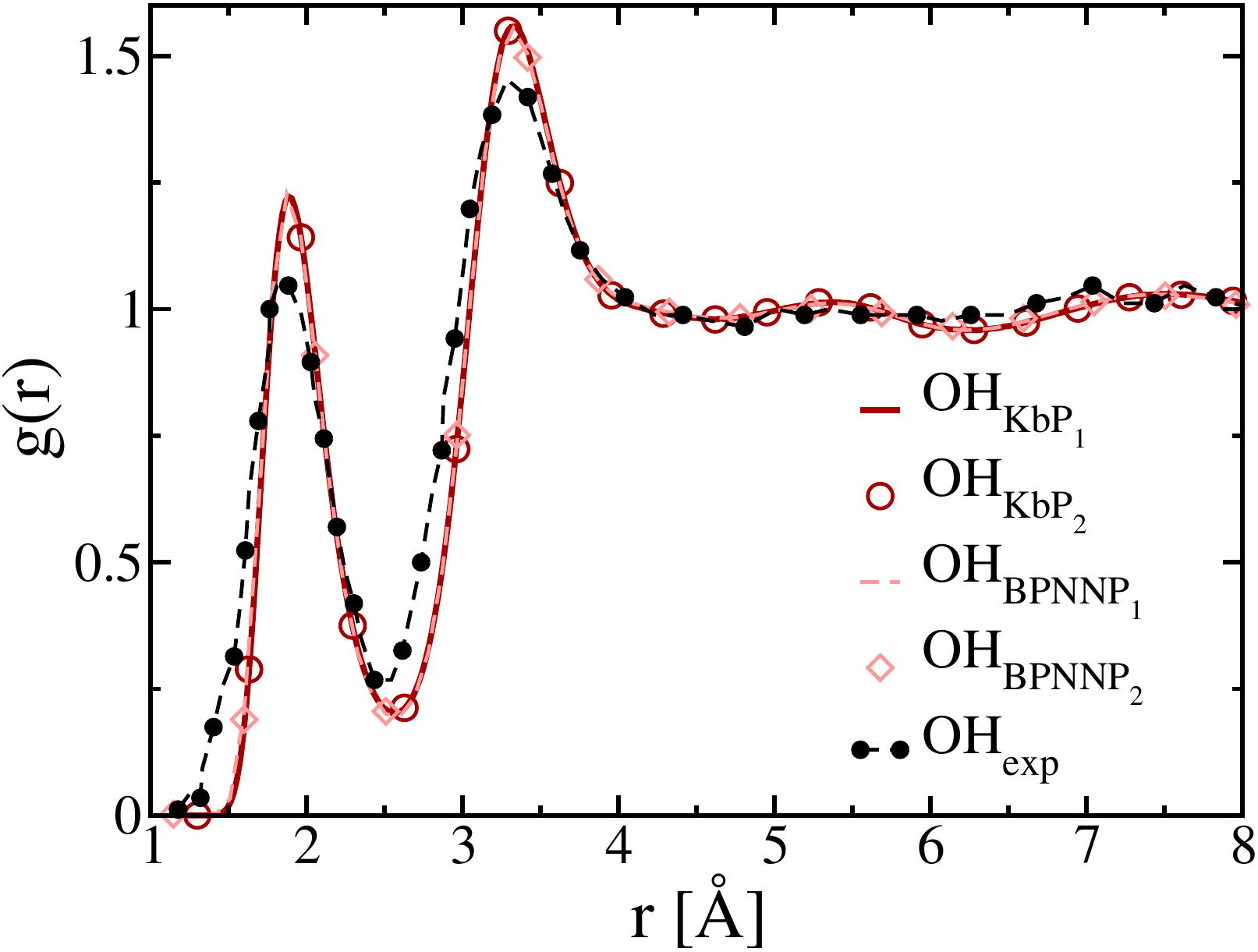} }
 \put(340,130){c)}
\end{picture}
\caption{\label{fig:rdf300} Radial distribution functions
 for KbP$_1$ (dark-solid lines), KbP$_2$ (solid circles), and BPNNP$_1$ (light-dashed lines and circles) at 300K for the RPBE+D3 functional.
 Experimental results~\cite{soper2013radial} are shown in black.
 Hydrogen-hydrogen pRDFs are shown in blue a), oxygen-oxygen pRDFs in green b), and oxygen-hydrogen pRDFs in red c).}
\end{figure*}

\begin{figure*}[htb]
\centering
  \begin{picture}(510,150)
 \put(0,0){\includegraphics[width=2.15in]{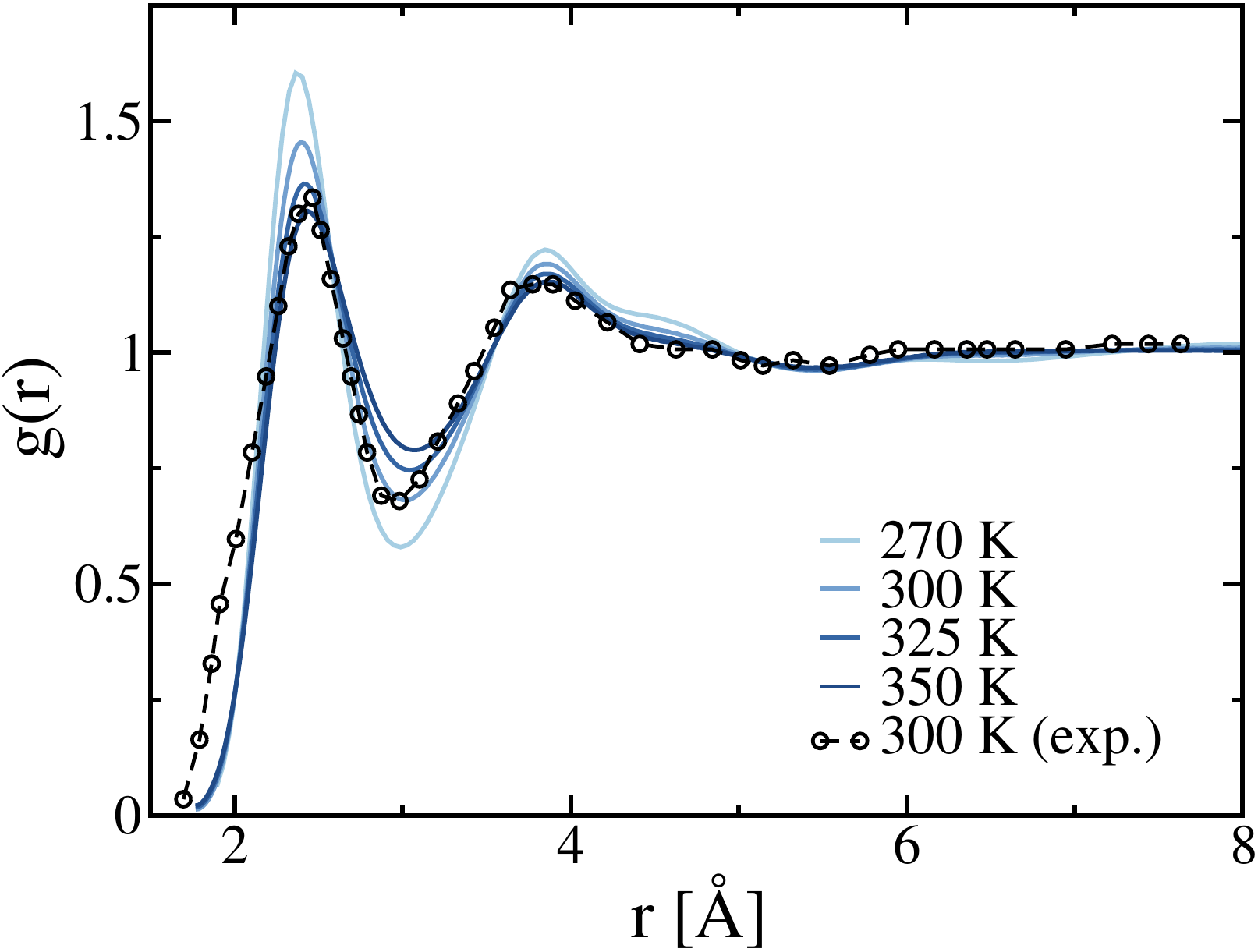} }
    \put(0,130){a)}
\put(170,0){\includegraphics[width=2.15in]{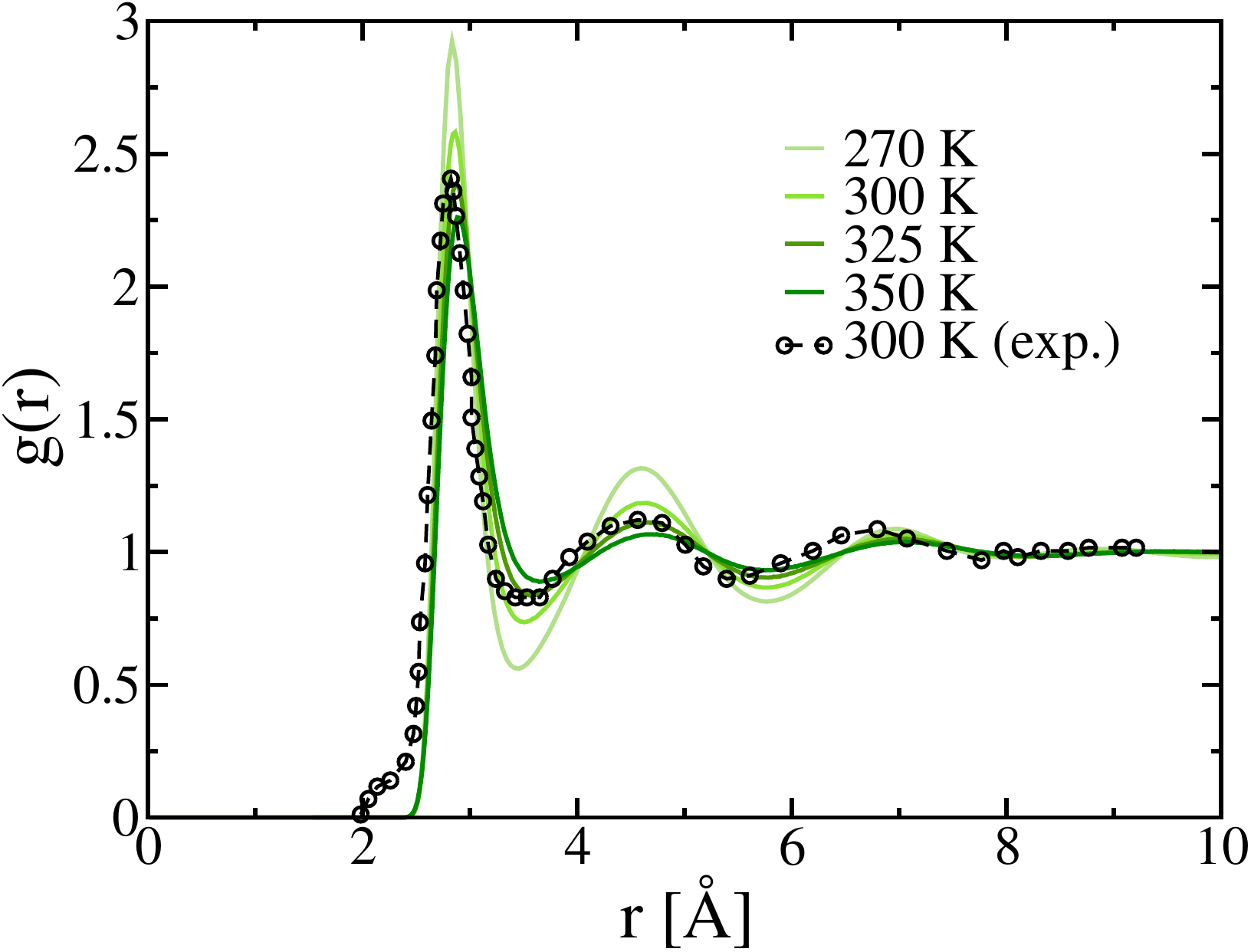} }
    \put(170,130){b)}
\put(340,0){\includegraphics[width=2.15in]{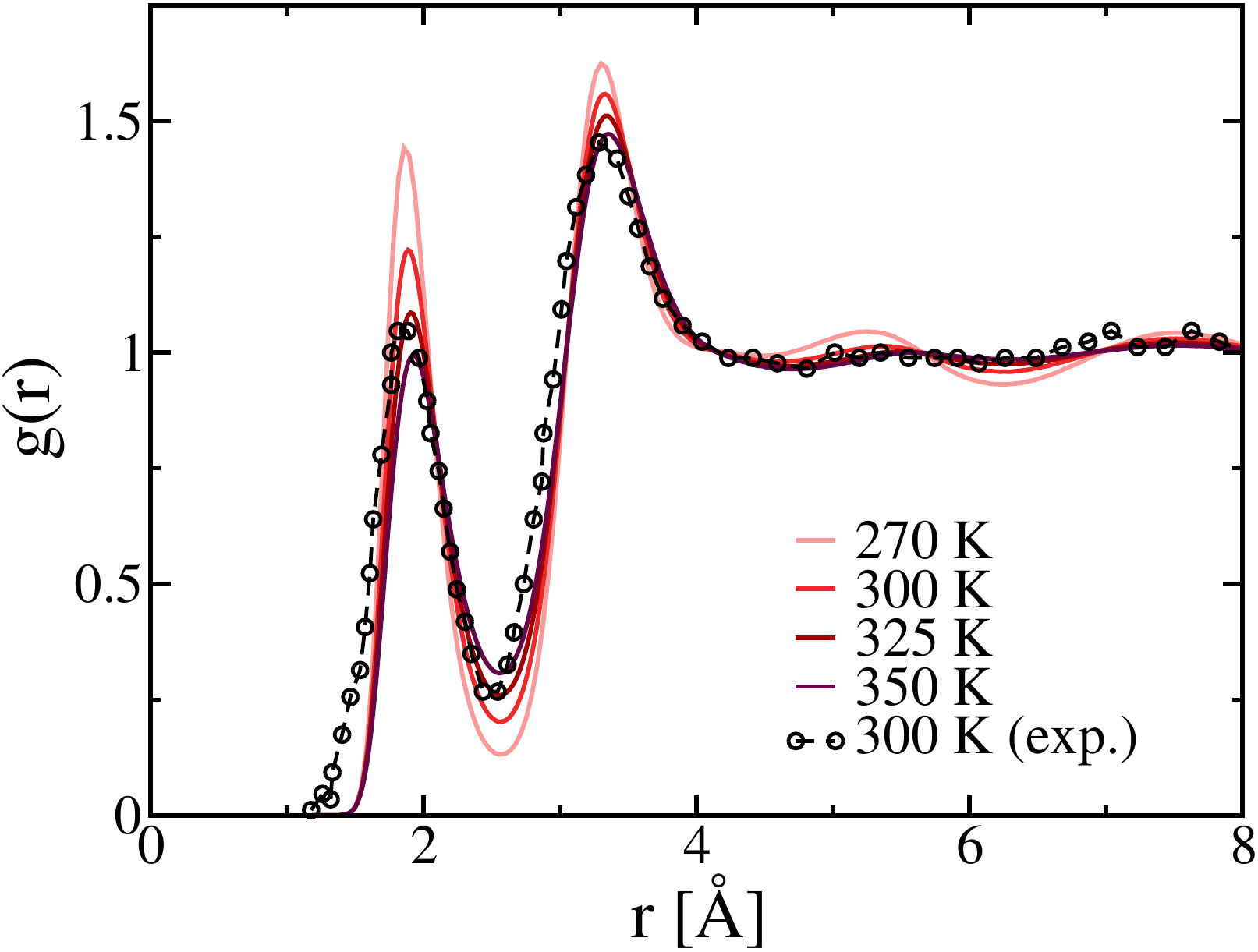} }
    \put(340,130){c)}
\end{picture}
\caption{\label{fig:rdfT} Effect of temperature on the partial radial distribution functions for a) hydrogen-hydrogen, b) oxygen-oxygen, and
 c) oxygen-hydrogen computed with the  KbP.}
\end{figure*}

Here, we compare the partial radial distribution functions (pRDF) for
 hydrogen-hydrogen (HH), oxygen-hydrogen (OH), and oxygen-oxygen (OO) at 300K.
 To do so, we run
 NpT simulations.
  In this case, the agreement between different MLPs is practically perfect, smaller than the line-width 
  as can be seen in Fig. \ref{fig:rdf300}.  Therefore, we decided to study
   the effect of temperature in the pRDF using only the KbP$_1$. 
    The different pRDFs are depicted in Fig. \ref{fig:rdfT}, for 270 K, 300 K, 325 K, 
    and 350 K. As expected, at lower temperatures water becomes more
     structured.

    To conclude this section, we comment on the comparison to experimental results. Clearly, agreement between the present simulations and experiment is excellent, in particular, considering that quantum effects are disregarded in the present work. The inclusion of quantum effects would lower and broaden the first peak in the pRDF. It is not uncommon to approximate the effect of quantum effects by increasing the temperature: indeed the pRDFs at 325 K improve upon the already good agreement at 300 K. \\

\subsection{Self-diffusion coefficient }

 Finally, we compare the MLPs for the self-diffusion coefficient $D$. To make sure that the dynamics is not affected by the use of a thermostat, we equilibrated the system
  first in the NpT ensemble and then performed a production run in the NVE ensemble using 3,546 molecules and a simulation time of 0.5 ns (1 million timesteps). The same volume and starting structures were used for all MLPs. As usual, the diffusion coefficient is determined from the mean square displacement using the relation: \\

     \begin{equation}
        \langle | \vec{r}(t) - \vec{r}(0)  |^2   \rangle = 6Dt.
        \label{eq:msd}
     \end{equation}

Again, the agreement between the methods is excellent, in particular considering the statistical uncertainties. The KbP, based on the larger dataset yields slightly larger self-diffusion coefficients than the BPNNP based on the same dataset. Generally, the self-diffusion coefficients are slightly lower than those obtained by Morawietz {\em et al.} \cite{morawietz2016van} and by Montero {\em et al.}\cite{montero2023kinetics}. 
Over the available temperature range, agreement with  experiments\cite{gillen1972self,holz2000temperature} is also excellent.

\begin{figure}[htb]
\centering
\includegraphics[width=3.1in]{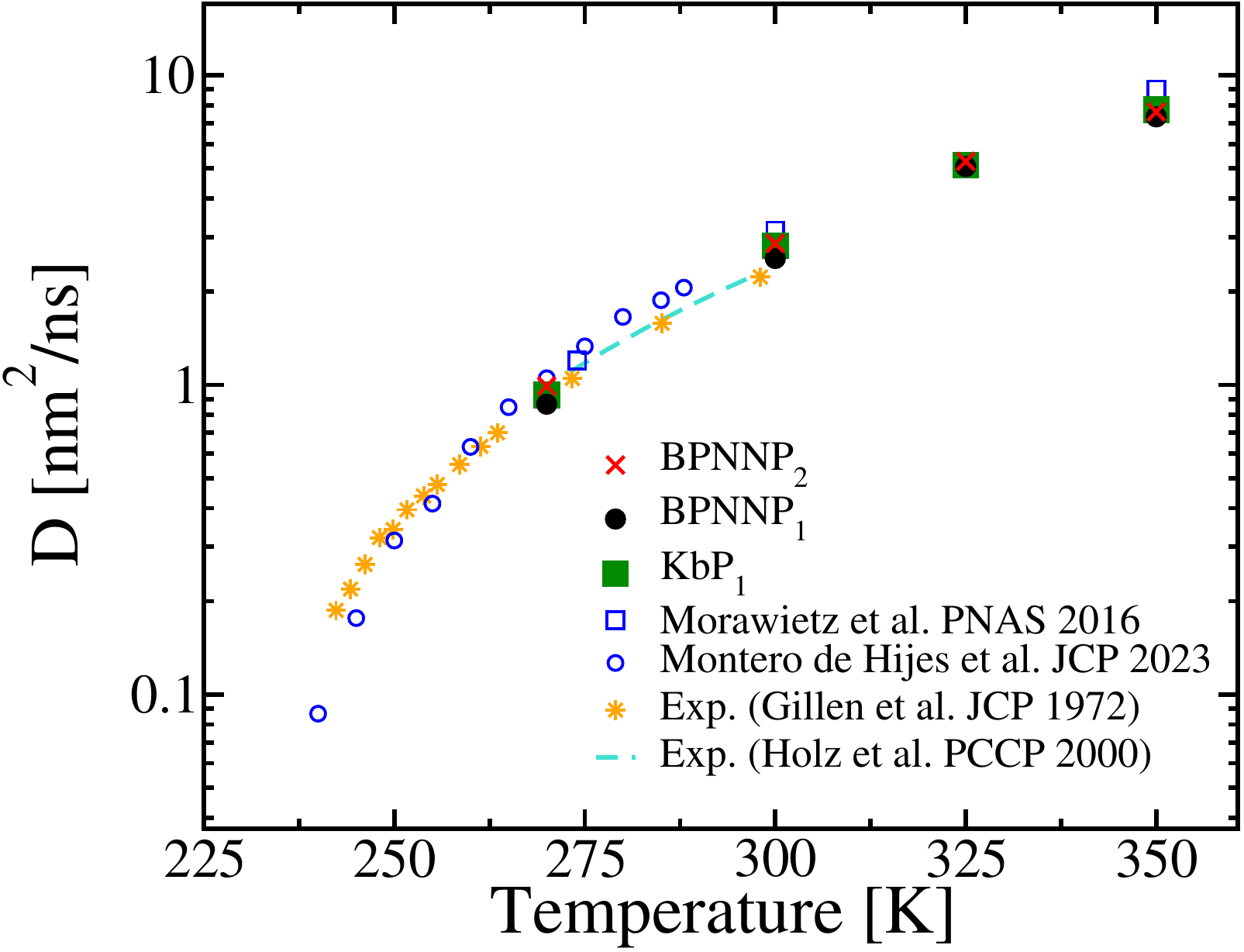} 
\caption{\label{fig:D} Self-diffusion coefficient $D$ as a function of temperature obtained from simulations with different MLPs. Also shown are experimental results.}
\end{figure}

\section{Discussion}

In this work, we have compared two well-established machine-learning approaches
 that permit to transfer the accuracy of first-principles calculations to molecular simulations approaching the cost of simple force-field models. These approaches 
 are the kernel-based method as implemented in VASP and the Behler-Parrinello neural network method implemented in n2p2. 
 For the first principles calculations, we use 
   RPBE+D3 with zero damping, which has previously  been 
   shown to reproduce 
   fairly well both structural and dynamical properties of water. We have taken great care to make our first-principles dataset technically as accurate as possible: for instance, increasing the k-point density yields negligible (i.e., sub meV/\AA) changes, and the cutoff was set to 2000~eV to minimize basis-set errors. Specifically, we emphasize that the use of a PW-cutoff of 1100 eV results in a Pulay stress error of about $-0.8$~kbar. This leads to an underestimation of the volume of about 5~\% at zero pressure, which is hardly acceptable.
      
We prepared two distinct first-principles datasets. Both start out from  450 structures obtained from on-the-fly learning by melting cubic ice, equilibrating just below the boiling point,  followed by several simulated annealing steps. The first set is extended by adding 1,000 structures from two parallel tempering runs covering the desired temperature range. The second set is based exclusively on on-the-fly learning adding 200 more structures generated again by simulated annealing. The first set is more diverse and expected to yield "better" MLPs.
     
For the KbP approach, we find that the results do not depend very significantly on the chosen database. Pair correlation functions are practically indistinguishable for the two KpP. 
Slight differences are found for the isobars at low temperatures where vitrification occurs. This however has no noticeable impact on the liquid state and, in particular, on the density maximum.
Results are less clear for the BPNNP. For the small training set, three out of four BPNNPs yield non-stable trajectories during parallel tempering, indicating that the neural network potentials require more training data and are more prone to instabilities when performing long simulations. For the large dataset, however, results from the BPNNP and the KbP simulations are very similar. Specifically, the radial distribution functions agree perfectly, within the line width of the shown plots. Also, there are no noteworthy differences in the self-diffusion coefficient. Furthermore, agreement with the experiment is very good for both BPNNP and KbP.

Despite the virtually perfect agreement for the pair correlation function and self-diffusion constant, slight but statistically significant differences are found for the density isobars. As mentioned, the KbP approach shows very consistent results for the two datasets as well as different settings. The BPNNPs show a more pronounced variation. When trained on the large dataset, we observe differences of about 1~\% in the predicted densities for four different MLPs. Furthermore, the average density is about 0.5~\% larger than for the KbP. The density maximum is also shifted slightly to higher temperatures. For the smaller training set --- as mentioned, only one out of four BPNNPs was stable --- the density for this single MLP was slightly higher than for the large database. It is not entirely trivial to explain this slight difference, but one possible explanation is that the BPNNP for technical reasons can not be trained on the pressure (to be precise the stress tensor). Recall that the pressure averaged over structures drawn from an NVT ensemble corresponds to the first derivative of the free energy. So if the pressure is not used for learning, potential errors for the equilibrium volume might be slightly larger. This in combination with the larger instability of the BPNNP potentials, might explain the slightly larger density that we obtain using BPNNPs. On the other hand, the predicted pressures using the BPNNP show no systematic error, so the previous arguments are not entirely conclusive. Unfortunately, we cannot even tell with certainty which of the two MLPs is more accurate. A full NPT first principles calculation at 2000~eV close to the density maximum is not feasible, as the dynamics of water is extremely slow.  We estimate that at least 0.2 ns (100,000 steps) first principles calculations would be required to make sufficiently accurate predictions on the density of water to determine which potential yields a result closer to the RPBE+D3 ground truth. Also reweighting of MLP-created ensembles using first-principles calculations did not yield a conclusive result.
 
\section{Summary and Conclusions}
\label{sec:conclusion}

In summary, the differences between the different MLPs are, in fact,  small. Specifically for the larger training dataset, the agreement between two entirely different MLPs is excellent, and {\em no difference} is noticeable for simple observables such as the diffusion constant or the pair-correlation function. 
The small discrepancies in the predicted density isobars are acceptable, at least for the large dataset. Recall that the difference in the density is only about 0.5~\% which translates into a lattice constant difference of only about $0.15~\%$. Such a difference is negligible, if we consider that the errors introduced by approximate density functionals are usually considered to be at least $0.5~\%$ for lattice constants, but often reach 3~\% for vdW or weakly interacting fragments. As a matter of fact, the density is underestimated by 10~\% using RPBE+D3.
The claim that the fitting errors are acceptable, is also supported by the observation that results obtained by fitting to an older dataset, based on the same density functional but computed with FHI AIMS, show a larger difference to the present work than the difference between the different MLPs fitted to the first-principles datasets obtained here. This indicates that the database is more relevant than how the database is fitted.

We finish with a word of caution. Root mean square errors for test datasets do not always convey the complete story. They are not indicative of stability, and even simple tests on say pair correlation functions might not be able to predict how large the differences might be for thermodynamic properties. It seems expedient to test at least two MLPs if some degree of certainty is desired, in particular, if it is not straightforward to simulate or check the observable using the full first-principles machinery.

\section{Acknowledgments}

The authors acknowledge the support from the  SFB TACO (project nr. F81-N) funded by Austrian Science Fund FWF as well as the computer resources and technical assistance provided by the Vienna Scientific Cluster (VSC).

\section{Author declarations}

\subsection{Conflict of Interest}

The authors have no conflicts to disclose. 



\bibliographystyle{ieeetr}
\bibliography{main}

\end{document}